\documentstyle[prl,aps,preprint,epsf]{revtex} \baselineskip 0.9truecm
\large  
\def\bel{\begin{equation} \label} \def\ee{\end{equation}}
\begin{document}
\draft \title{Non-universal exponents in interface growth} \author{T.
  J. Newman$^{1}$ and Michael R. Swift$^{2}$}
\address{$^{1}$Department of Theoretical Physics, University of
  Manchester,\\ Manchester, M13 9PL, UK\\ $^{2}$International School
  for Advanced Studies,\\ via Beirut 2-4 and Sezione INFM di Trieste,
  I-34013} \maketitle
\begin{abstract}
  We report on an extensive numerical investigation of the
  Kardar-Parisi-Zhang equation describing non-equilibrium interfaces.
  Attention is paid to the dependence of the growth exponent $\beta $
  on the details of the distribution of noise $p(\xi)$.  All
  distributions considered are delta-correlated in space and time, and
  have finite cumulants. We find that $\beta $ becomes progressively
  more sensitive to details of the distribution with increasing
  dimensionality. We discuss the implications of these results for the
  universality hypothesis.
\end{abstract}
\vspace{5mm} \pacs{PACS numbers: 05.40.+j, 68.35.R}

\newpage The Kardar-Parsis-Zhang (KPZ) equation\cite{kpz,rev1,rev2},
although originally introduced as a model of non-equilibrium interface
growth, has acquired a broader significance over the past decade as
one of the simpler examples of a strong-coupling system. To this date
there exists no systematic analytic procedure for determining the
properties of the model for large values of the non-linear coupling.
Additional interest in the model stems from its intimate mathematical
relation to two other important systems, namely the noisy Burgers
equation\cite{burg}, and the equilibrium properties of a directed
polymer in a random medium (DPRM)\cite{rev2}.

Most work on this problem to date has concentrated on determining the
properties of the system (e.g. values of dynamic exponents) in the
strong-coupling phase. In pursuit of this goal, the `hypothesis of
universality' (HOU) has been generally adopted; namely that details of
the model representation of the KPZ equation should not affect
universal quantities such as exponent values. In this Letter we report
on an extensive numerical investigation of the KPZ equation which
shows that exponents are sensitive to the precise form of the noise
distribution, this sensitivity becoming extreme in higher dimensions.
We regard our results as being {\it strongly suggestive} of a
breakdown of universality in the KPZ equation, but of course, they
cannot constitute a proof of this assertion.  To set the scene for
what is to come we devote one paragraph below to the general
theoretical framework of the KPZ equation, and a further paragraph to
the current state of knowledge.

Denoting the interface profile by $h({\bf x},t)$ (which is defined
perpendicular to the $d$-dimensional substrate), the KPZ equation has
the form
\begin{equation}
\label{kpz}
\partial_t h = \nu \nabla ^{2}h + \lambda (\nabla h)^{2} + \xi \ ,
\end{equation}
where $\xi ({\bf x},t)$ is a stochastic source, generally taken to be
delta-correlated in space and time: $P[\xi ] = \prod _{\bf x} \prod
_{t} p(\xi ({\bf x},t)) $. The canonical choice for the distribution
$p$ is a gaussian, and the HOU is invoked to argue that any other
choice (so long as it has finite cumulants) will lead to the same
large-scale behavior.  The prime goal is the computation of the
exponents which characterize the evolution of fluctuations, along with
the value of the upper critical dimension $d_{u}$, above which one
expects the exponents to saturate at their mean-field values. In the
language of the KPZ equation, there are three exponents of immediate
interest: Starting from a flat interface, the mean square fluctuations
grow as $W^{2} \equiv \langle h^{2} \rangle _{c} \sim t^{2\beta}$, but
will saturate for large times in a finite system of linear dimension
$L$ -- the saturated value is expected to scale as $W \sim L^{\chi }$.
The fluctuations over the entire temporal regime may be conveniently
described by the two-point correlation function $C({\bf r},t) \equiv
\langle (h({\bf r},t)-h({\bf 0},t)) ^{2}\rangle \sim r^{2\chi}
f(r^{z}/t)$. The exponent $z$ is often known as the dynamic exponent
and gives the fundamental scaling between length and time.  The three
exponents may be reduced to one independent exponent by two scaling
laws: $z\beta = \chi$ and $z + \chi = 2$. The first law comes from the
scaling behavior of $C$ as $r \rightarrow \infty$, and the second is a
consequence of invariance of the equation under an infinitesimal tilt
transformation.  It is also worth mentioning that the non-linear
transformation $w=\exp(\lambda h/\nu)$ produces the linear equation
\begin{equation}
\label{linear}
\partial _{t} w = \nu \nabla^{2}w + (\lambda /\nu)w \ \xi \ ,
\end{equation}
which corresponds to the equation for a $(d+1)$-dimensional directed
polymer in a random potential $-\lambda \xi$ .

Given the extreme difficulty of any analytic progress, most
investigations of the KPZ equation have been numerically based. These
investigations fall into two categories. The majority of numerical
work consists in simulating microscopic models which under the HOU are
assumed to have the same large-scale behavior as the KPZ equation.
Such models include Eden growth\cite{eden}, polynuclear growth
models\cite{png}, the restricted solid-on-solid (RSOS)
model\cite{rsos,an}, and its close relative, the hypercube stacking
model\cite{hsm}. The second numerical approach is that of direct
integration of the equation itself\cite{af,wolf}.  This entails some
subtleties of discretization which have only recently come to
light\cite{nb}. However, there has been reasonable agreement between
these various approaches as regards the values of exponents, at least
in $d=1+1$ (where there exists an exact analytic result of
$z=3/2$\cite{exact}) and $d=2+1$.  In higher dimensions, numerical
work becomes more difficult due to the smaller linear dimension of the
systems (and the consequent early onset of finite size effects), but
still there has been general agreement that although exponents may not
be so precisely determined, there is no sign of a crossover to mean
field values ($z=2$) for dimensions up to $d=7+1$\cite{an}.  An
interesting analytic method which may be applied to the strong
coupling regime is mode-coupling theory\cite{mc} (although it is based
on an {\it ad hoc} neglect of vertex renormalization) which seems to
support $d_{u}=4$. Also there has been recent work\cite{lk} based on
short-distance expansion techniques in the renormalization group (RG),
which indicates the bound $d_{u} \le 4$, whereas a mapping to directed
percolation\cite{maj} suggests that $d_{u} \le 5$.

Our original motivation for this numerical study was to integrate a
recently proposed discrete equation\cite{nb}, which was shown to
correctly capture the strong-coupling behavior of the KPZ equation on
a lattice. Only during the course of our work did we discover the
noise sensitivity of the exponents, which is the focus of this Letter.
Nevertheless, before presenting our results, it is useful to briefly
describe this improved algorithm, and also to exhibit its close
relation to the algorithm for zero temperature DPRM\cite{rev2}, which
has also been numerically studied in the past\cite{kbm}.

The key point concerning the discretization of the KPZ equation, is
that one is only guaranteed to capture the strong-coupling physics by
discretizing the directed polymer equation (\ref{linear}), and
constructing the discrete KPZ equation by the inverse transform
$h_{i}=(\nu/\lambda)\ln(w_{i})$.  The time discretization necessarily
introduces a two-stage process -- i) pumping with the noise, and ii)
relaxing with the deterministic part of the equation. Explicitly one
has the discrete KPZ equation in the form
\begin{eqnarray}
\label{discrete}
\nonumber {\tilde h}_{i}(t) & = & h_{i}(t) + \Delta ^{1/2} \xi _{i}(t)
\\ h_{i}(t+\Delta) & = & {\tilde h}_{i}(t) + (\nu /\lambda)\ln \left
\lbrace 1 + (\Delta \nu / a^{2}) \sum \limits _{j \ nn \ i} \left
\lbrack e^{\lambda ({\tilde h}_{j}-{\tilde h}_{i})/\nu } - 1 \right
\rbrack \right \rbrace \ ,
\end{eqnarray}
where $\Delta$ and $a$ are the grid scales for time and space
respectively.  Taking the strong-coupling limit $\lambda \rightarrow
\infty$ allows one to write the relaxation stage of the above equation
in the much simpler form
\begin{equation}
\label{sc}
h_{i} (t+\Delta) = \max \limits _{j \ nn \ i} \left ( {\tilde h}_{i},
\lbrace {\tilde h}_{j} \rbrace \right ) \ .
\end{equation}
This corresponds exactly to the zero-temperature DPRM algorithm but
written here in terms of the field $h_{i}$ rather than $w_{i}$. Note
that there are no adjustable parameters in the strong-coupling
algorithm, except for the functional form of the noise distribution
$p(\xi)$ (since the grid scales, and the noise strength may be scaled
away).

We have implemented the above algorithm in dimensions $d+1$ with
$d=2,3,4$, with a flat surface ($h_{i}(0)=0$) as the initial
configuration..  The only adjustable `parameter' in our simulations
has been the function $p(\xi)$. This function was taken to be either
gaussian $p_{g}(\xi ) \sim \exp (-\xi^{2}/2)$, or of the form
$p_{\alpha}(\xi) \sim (\sigma -|\xi|)^{\alpha}, \ -\sigma \le \xi \le
\sigma$ (where $\sigma $ is adjusted to maintain unit variance for
each choice of $\alpha $). On varying $\alpha $ the distribution
$p_{\alpha }$ interpolates through the forms: bimodal ($\alpha
\searrow -1$), top-hat ($\alpha = 0$), triangular ($\alpha = 1$), and
finally distributions with very rapidly vanishing tails as $\alpha
\rightarrow \infty$. Note that all distributions are strongly
localized and have an infinite set of finite cumulants. [They are thus
distinct from distributions with {\it power-law tails}, which are
known to change the exponent values\cite{rev2}.]

Our simulations are performed on lattices of size $L^{d}$, with
averaging over $N$ samples. The simulations were of the size
$(L=2048,N=8)$ for $d=2$, $(L=200,N=8)$ for $d=3$, and $(L=60,N=8)$
for $d=4$. Although these simulations are of the largest scale
possible within our resources (a cluster of DEC alpha workstations),
we are aware that specialist computational groups can improve on the
precision of our results. However, the qualitative features of
interest in this Letter are convincingly clear from our simulations.

The interface width $W$ is plotted as a function of time in Figs.1, 2
and 3, for dimensions $d=2,3$ and 4 respectively. As mentioned
earlier, this quantity is expected to grow as $W \sim t^{\beta}$ (so
long as the dynamic length scale is much less than $L$).  It is clear
that there is a dependence of $\beta $ on the value of $\alpha $, this
dependence becoming progressively stronger in higher dimensions. (This
is the reason for not presenting data in dimension (1+1), for which
all distributions yield a value of $\beta $ close to the exact result
of 1/3\cite{exact}).  The measured values of $\beta $ are presented in
Table 1.  An important point to make is that in all dimensions, our
results are in agreement with the consensus of exponent values in the
literature\cite{an,hsm,kbm}, {\it if} we use the gaussian noise
$p_{g}(\xi)$.  However, the values of $\beta$ drops smoothly as the
parameter $\alpha $ is decreased.

Naturally one may attempt to interpret these results in terms of a
temporal crossover. Given the straightness of the curves, such a
crossover would be exceptionally slow. We have checked for the
presence of this phenomenon by trying to fit the data with the generic
form $W^{2} = At^{2\beta} + Bt^{2\gamma}$ (which implicitly includes
the popular fitting Ansatz $\gamma=0$ used by previous
groups\cite{hsm,kbm}). The best fits correspond to values of $\beta $
close to the value one would obtain by the simpler fit $W \sim
t^{\beta}$, with the `correction to scaling' exponent $\gamma $ taking
a value of approximately $\beta /2$. This indicates that the curves
show no sign of crossover, as they have a clearly dominant power-law
form, with modest corrections. The quoted errors in $\beta $ (shown in
Table 1) are estimated from the range of $\beta $ with which one may
obtain an acceptable fit using the above fitting Ansatz.

To enable a check of our results, we have concentrated on the value
$\alpha =1/2$ in $d=3$ for which we averaged over $N=64$ samples of
size $L=180$. In this case, the data is of a good enough quality to
measure a running value of the growth exponent $\beta _{\rm eff}(t) =
d\ln(W)/d\ln(t)$\cite{kbm}. The quality of the data obtained for the
value $\alpha=-1/2$ (with $N=8$ samples) is also sufficiently good to
allow this measurement.  The running exponents $\beta _{\rm eff}(t)$
for $\alpha=1/2$ and $\alpha=-1/2$, plotted against $t^{-\beta}$
(which magnifies potential systematic deviations from a simple power
law), are shown in the insets of Figs. 4 and 5 respectively; with the
previously measured values of $\beta $ used in each case.  There is no
sign of any asymptotic deviation away from these values of $\beta =
0.14$ and $\beta = 0.05$.

Another independent check was made for these two values of $\alpha $,
by studying the variation of the saturated (or steady-state) width
$W_{SS}$, as a function of the system size. On scaling grounds one
expects $W_{SS} \sim L^{\chi}$, with $\chi = 2\beta/(1+\beta)$,
where we have assumed the exponent relations 
$z\beta = \chi$ and $z + \chi = 2$.
In Figs.4 and 5, we plot $W_{SS}$ against $L$ for $\alpha=1/2$ and
$\alpha=-1/2$ respectively. One observes that there is more than a decade
of clean scaling in each case, with fitted values $\chi = 0.24$ and
$\chi = 0.08$ respectively, which are consistent (within the stated
errors) with the previously measured values of $\beta $.  On the
grounds of this numerical work we are led to the statement that the
growth exponent $\beta$ is non-universal with respect to changes in
the noise distribution $p(\xi)$. In the remainder of the paper we
present a brief discussion of the robustness and physical implications
of this claim.

It is possible to gain information about the exponents from physical
quantities other than $W$. As an example, one may measure the
two-point correlator $C({\bf r},t)$ (or its Fourier transform, the
structure factor) and attempt to collapse the functions using
dynamical scaling. We have been able to produce tolerably good data
collapse for $C({\bf r},t)$ in all cases, yielding values of the
exponents within the errors of those listed in Table 1. However, in
our opinion the measurement of exponents from data collapse is less
reliable than direct measurements of $W$ due to the difficulties of
including corrections to scaling in a systematic fashion.

As mentioned above, one may try to interpret these results in terms of
slow temporal crossover. Our analysis indicates that strong crossover
effects are absent from the data, since we are able to measure
reasonably precise values of the exponent $\beta $ along with its
`correction to scaling exponent' $\gamma$.  However, one can not rule
out the possibility that one has simply failed to reach the `true'
asymptotic regime (AR), and that one is measuring some transient
scaling regime. This introduces the question of how one should
empirically define the AR.  For our purposes we have used the standard
working definition that in the AR one observes clean power-law
behavior of the quantities of interest (in our case the interface
width $W$), and that the correlation length in the system is large
compared to the lattice scale. Both criteria have been met in our
simulations.  A further criterion which one may invoke\cite{krugpc} is
that the interface width itself must be much greater than the
effective lattice spacing in the growth direction (here set to $O(1)$
by the noise variance being normalized to unity.) This criterion is
not satisfied in our simulations for $d=4$, or for very small
$\alpha$, for the simple reason that if the interface fluctuations
grow very slowly with time (meaning $\beta $ is small) then one will
never achieve $W \gg 1$ on observable time-scales, even though the
correlation length of the system is large (since $z$ is generally
close to 2).  There can be no definitive answer to the question of
whether a given simulation has reached the AR, although we would
encourage specialist computational groups to improve on our time
scales and system sizes in order to shed more light on this question.

As a final comment on the simulations, it is possible that these results 
are a consequence of working at the strong-coupling limit ($\lambda
\rightarrow \infty$). An example is known in the directed polymer
literature\cite{khh} of exponents dependent on the noise distribution
at zero temperature -- for the case of perfectly correlated disorder
in the longitudinal direction. The reasons for this dependency are
physically clear and may be illustrated by a simple Flory argument. 
However, it
is doubtful whether a similar effect is relevant to our simulations,
in which the noise is delta-correlated in time.

If the HOU is false for the KPZ equation, it is crucial to trace the
underlying physical reasons for this. The most fragile aspect of KPZ
physics is the role of the microscopic cut-off. For instance, the fact
that the naive (but canonical) discretization fails to retain the
continuum physics\cite{nb} gives one real cause for concern.  It may
be that the lattice scale is always relevant to the large-distance
scaling of the interface, which would then give scope for
non-universal features, such as those seen in the present work. A
similar example of non-standard scaling is known from controlled
calculations on the deterministic Burgers problem\cite{esnew}.

In this paper we have given convincing numerical evidence that the KPZ
growth exponent $\beta $ is strongly sensitive to certain details of
the noise distribution (here characterized by the parameter $\alpha
$).  It is important to understand whether the results presented here
are really the hallmark of a breakdown of the HOU for the KPZ
equation, or whether there exist (exponentially) long crossover scales
deep within the strong-coupling phase itself. Whichever is the case,
we believe that these results are indicative of unexpected and
interesting new physics in the KPZ problem.

\vspace{5mm}

The authors are grateful to Alan Bray, Joachim Krug, Amos Maritan and
Michael Moore for useful conversations.  TJN acknowledges financial
support from the Engineering and Physical Sciences Research Council.

\newpage
\noindent
{\bf List of figure captions}

\vskip 1.in
\noindent
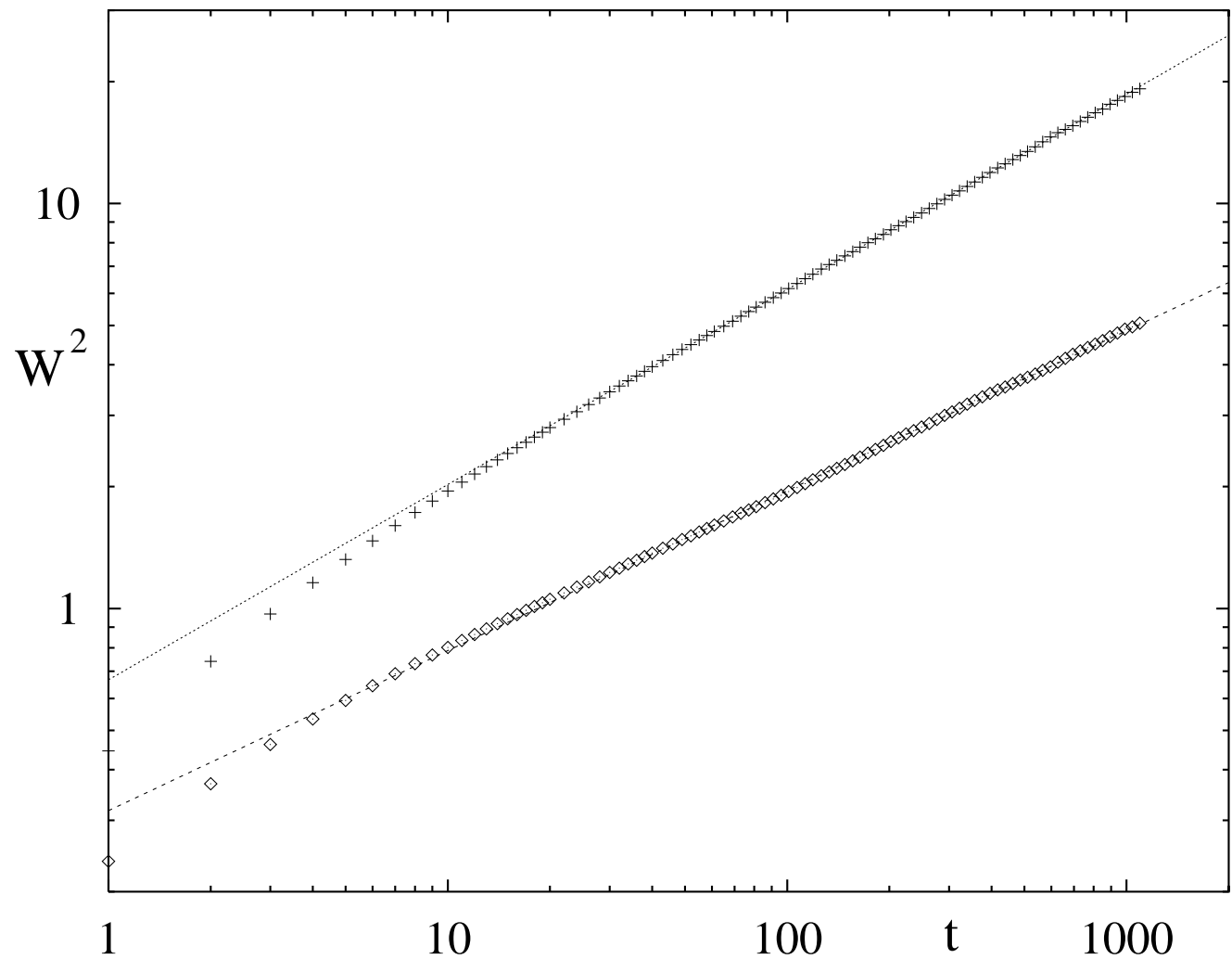
Fig.~1: Interface width $W^{2}$ versus $t$ in dimension $2+1$.
The upper curve corresponds to $p_{g}$ and the lower
to $p_{\alpha}$ with $\alpha=0$. The straight lines are
fitted with values of $\beta $ given in Table 1.

\vskip .5in
\noindent
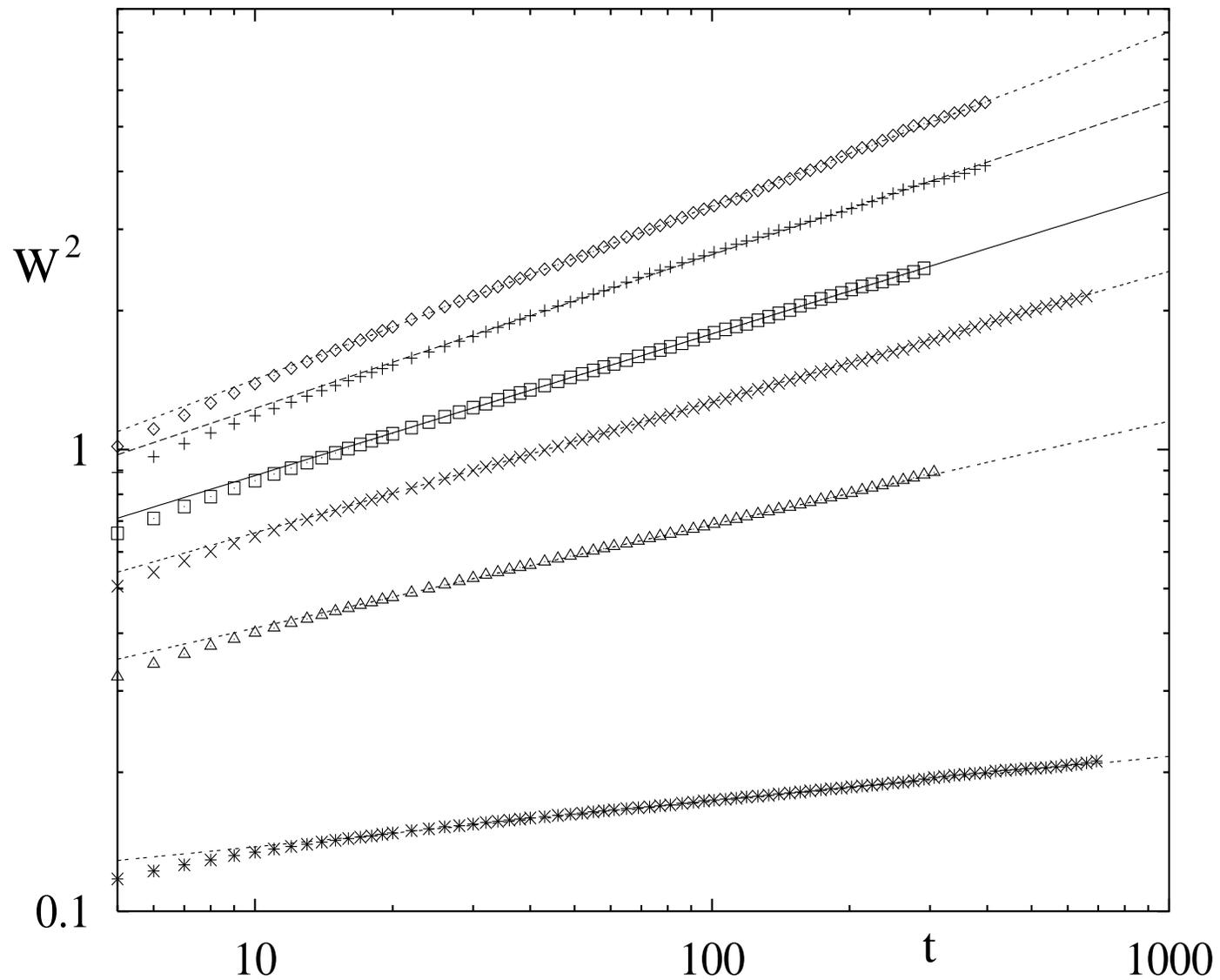
Fig.~2: Interface width $W^{2}$ versus $t$ in dimension $3+1$.
The uppermost curve corresponds to $p_{g}$ and the lower curves
to $p_{\alpha}$ with $\alpha=2,\ 1,\ 1/2,\ 0$ and $-1/2$ in descending
order. 

\vskip .5in
\noindent
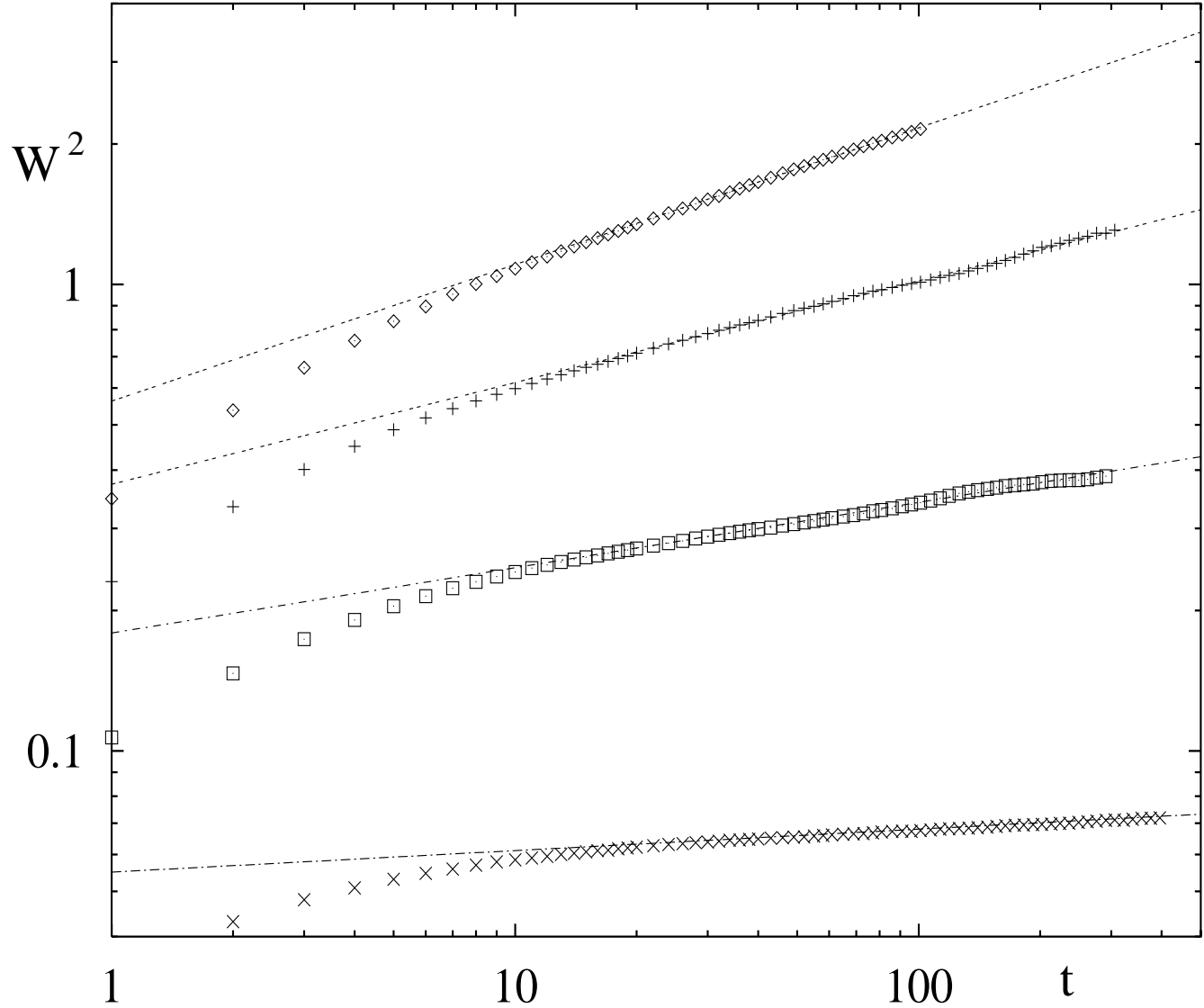
Fig.~3: Interface width $W^{2}$ versus $t$ in dimension $4+1$.
The uppermost curve corresponds to $p_{g}$ and the lower curves
to $p_{\alpha}$ with $\alpha=1,\ 0,\ -1/2$ in descending order.

\vskip .5in
\noindent
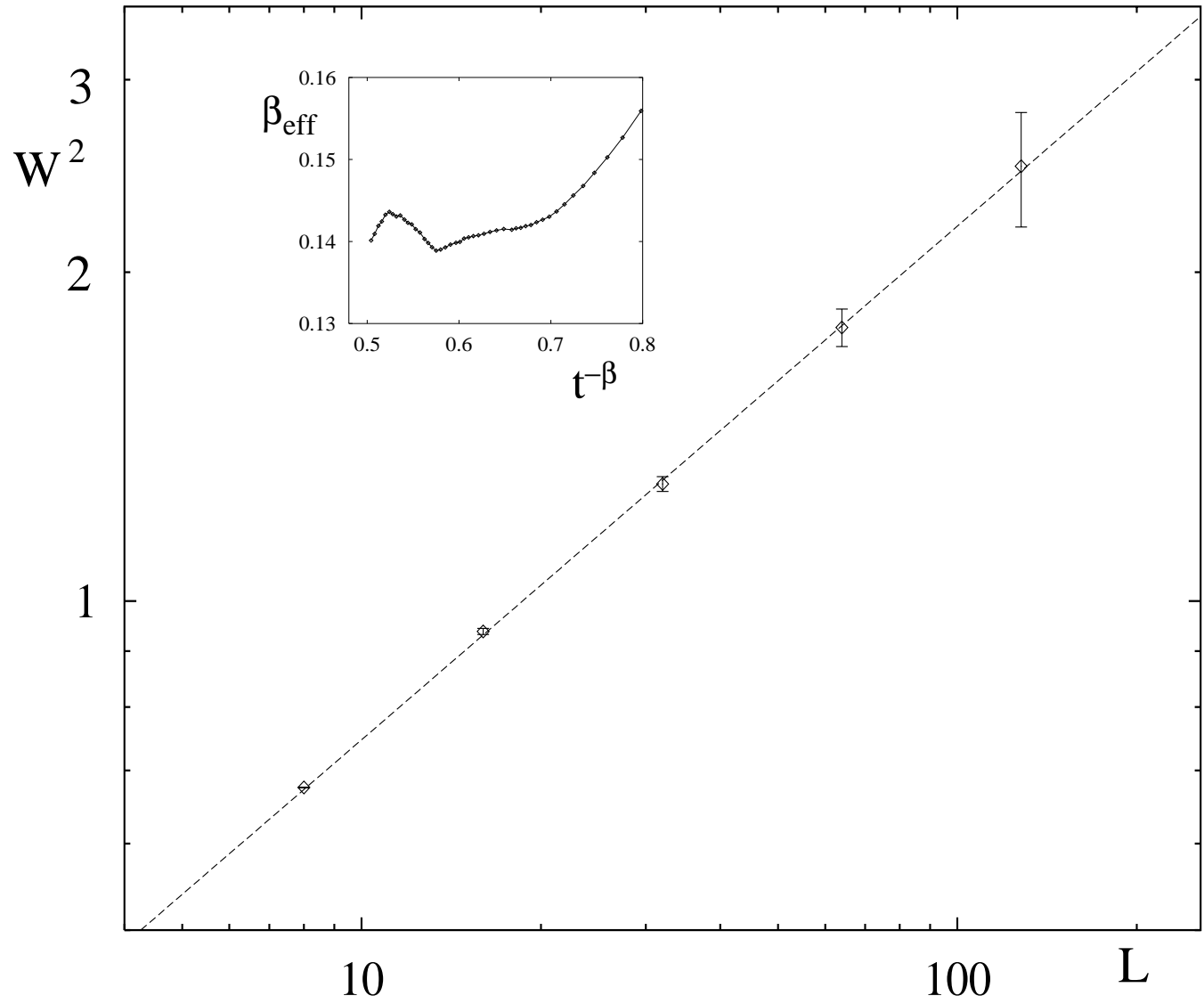
Fig.~4:  Steady-state interface width $W^{2}_{SS}$ versus system
size $L$ in dimension $3+1$, using distribution $p_{\alpha }$
with $\alpha = 1/2$. The fitted line has a slope of $2\chi = 0.48$.
The inset shows the running exponent $\beta _{\rm eff}(t)$ versus 
$t^{-\beta}$ with $\beta=0.14$.

\vskip .5in
\noindent
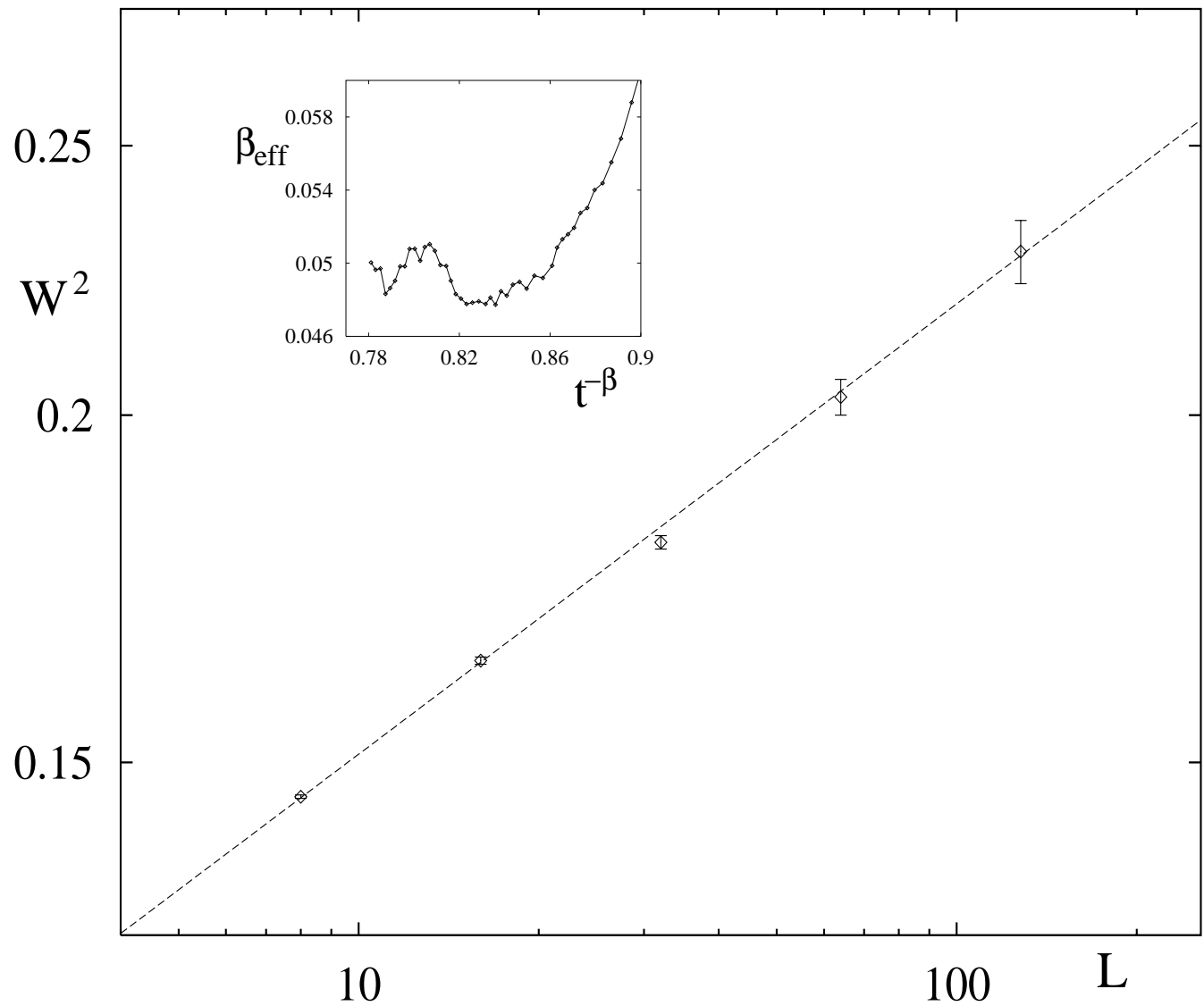
Fig.~5: Steady-state interface width $W^{2}_{SS}$ versus system
size $L$ in dimension $3+1$, using distribution $p_{\alpha }$
with $\alpha = -1/2$. The fitted line has a slope of $2\chi = 0.16$.
The inset shows the running exponent $\beta _{\rm eff}(t)$ versus 
$t^{-\beta}$ with $\beta=0.05$.

\vskip .5in
\noindent
Table.~1: Values of the measured growth exponent $\beta $ as
a function of dimension $d$, and noise distribution $p(\xi)$.

\newpage

\begin{center}
{\bf Table.~1}

\vskip .7in

\begin{tabular}{|c||c|c|c|c|c|c|}\hline
  \raisebox{-1.5ex}[1.5ex] {$d+1$} 
  & \multicolumn{5}{|c|}{$p_{\alpha}$} & \raisebox{-1.5ex}[1.5ex] 
   \ \ \ \ {$p_{g}$} \ \ \ \ \\ 
   \cline{2-6} & $\alpha=-1/2$ & \ \ $\alpha=0$ \ \ & $ \ \alpha=1/2$ \ & 
   $ \ \ \alpha=1$ \ \  & \ \ $\alpha=2$ \ \ & \\ \hline \hline 

  2+1 & -- & 0.19(1) & -- & -- & -- & 0.24(1) \\ \hline

  3+1 & 0.05(1) & 0.11(1) & 0.14(1) & 0.15(1) & 0.165(10) & 0.185(10) \\ 
\hline

  4+1 & 0.025(10) & 0.07(1) & -- & 0.11(1) & -- & 0.14(1) \\ \hline 
\end{tabular}

\end{center}

\end{document}